\documentclass[11pt]{article}
 \usepackage{geometry} 
 \usepackage{graphicx}
\usepackage{hyperref}
 \usepackage[symbol]{footmisc}

  \usepackage{ulem}
 \usepackage{setspace,ifdraft,blindtext}

\title{From gamma-ray bursts to fast radio bursts}
 \author{S.\ R.\ Kulkarni\\
 California Institute of Technology, Pasadena, CA 91125, USA}
\date{} 

\begin{document}
 \maketitle

Gamma-ray bursts (GRBs) were discovered  in late sixties by a team
based at the Los Alamos National Laboratory but reported$^1$ in
June, 1973.  Fast radio bursts (FRBs) were reported about a decade
ago.$^2$  GRBs and FRBs share two letters in their acronym, one
common word and perhaps a similar history.  Here, I review  the
development of the GRB field and use this occasion to apply lessons
learnt during the early days of GRB  to the rapidly developing
field of FRBs.

Gamma-ray ($\gamma$-ray) detectors have a large field-of-view but
lack the ability to make images. The lack of localization meant
that the distance to the bursts was unconstrained which, in turn,
led to a frenzy of speculation. At the Seventh Texas Symposium
(Dallas, December 1974) the Los Alamos  team reported a total of
33 GRBs.  Malvin Ruderman$^3$ summarized the constraints posed by
the observations to theories of GRBs and also listed the many more
proposed models.  The prevailing view was that  GRBs were nearby
old Galactic neutron stars.

The proceedings$^4$ of a conference held at Taos (July, 1990) at
the eve of the launch of the Compton Gamma-ray Observatory ({\it
CGRO}) make for an excellent reading.  The neutron star model rested
on the presence of  absorption lines in  some GRBs. These were
interpreted as cyclotron lines arising in the photosphere of neutron
stars.  The isotropic distribution meant that the neutron stars
were nearby, distance $\approx 100\,$pc.  The resulting energetics
were modest.  Triangulation of bright events, based on the arrival
time provided by the Inter-Planetary Network (IPN) of satellites,
led to localizations of a few square arc-minutes.  The absence of
optical quiescent counterparts was consistent with a neutron star
origin.  In any case, the Burst and Transient Source Experiment
(BATSE) aboard {\it CGRO} was expected to make definitive measurements
of the sky distribution of faint events.

To be fair, suggestions for extra-galactic models were made in the
eighties.  In an influential paper$^5$, Bohdan Paczy\'nski, noting
the isotropic distribution and  the slow rise in the counts of
fainter sources, advocated an extra-galactic model. He noted that
the inferred isotropic $\gamma$-ray energy release was $10^{51}\,$erg
-- similar to the mechanical  energy yield of supernovae.  Paczy\'nski
strengthened his case by attributing three repeating bursts to an
extragalactic GRB lensed by an intervening galaxy.  However, no
physical mechanism  which could produce short bursts of $\gamma$-rays
with high efficiency was identified.

By the time of the 1995 seminal review$^6$ by Gerald Fishman and
Charles Meegan,  two classes had emerged: short duration with  hard
spectrum (SHB) and long duration but with soft spectrum (LSB).  A
subset, the eponymously named soft $\gamma$-ray repeaters (SGR)
were accepted to be of Galactic origin (with the prototype, ``5
March 1979" residing in the Large Magellanic Cloud).  The three
bursts discussed above had been identified with SGR\,1900+14.  Next,
despite the increased sensitivity of BATSE, no absorption features
were seen.  Finally, the faint bursts were isotropically distributed
which, in the Galactic framework, required the neutron stars to be
located in the  Galactic ``corona"  (distance $\approx 100$\,kpc).
The introductory statement -- ``...  adopt a new paradigm for the
origin of $\gamma$-ray bursts: sources at cosmological distances.
This new paradigm is almost entirely by default" -- and the closing
statement of the review -- ``The $\gamma$-ray enigma appears to be
as great now as it was 20 years ago" -- summarizes the mood of that
era.

Another highlight of 1995 was the ``The Great Debate" featuring
Paczy\'nski arguing for a cosmological origin and Donald Lamb for
a Galactic origin. See \url{https://apod.nasa.gov/debate/debate95.html}
for the program, presentations and photographs. I was present at
this debate. Upon entry to the auditorium, attendees, based on their
conviction, were expected to pick and wear one of three buttons:
``Galactic", ``Extragalactic" and ``Other".  I collected 'em all!
There was no winner.  The Galactic  model was  strained, requiring
a goodly fraction of high velocity neutron stars with tailor-made
bursting duty cycle. The extra-galactic model  failed to explain
the absorption lines.  Both speakers advocated for a mission to
search for GRBs towards our neighboring galaxy (M31).

Theoretical studies of extra-galactic origins accelerated.  In an
influential conference summary, Stanford Woosley$^7$ discussed
neutron star quakes, neutron phase transitions and the ``collapsar"
model (in which a natal black hole accretes matter copiously). Many
focused on neutron star coalescences$^8$ while others investigated
long-lived radiation from GRB ``remnants"$^9$.  Motivated thus,
sensitive ground-based searches for the remnants accelerated. Shortly
thereafter, {\it BeppoSAX} discovered$^{10}$ decaying X-ray emission
from GRB\,970228.  The ``afterglow'' phenomenon -- long lived (hours
to months) emission at lower energies (X-rays through radio) --
enabled arcsecond localization. This quickly led to a demonstration
of the extra-galactic, nay, cosmological ($z\approx 0.8$), origin$^{11}$
of LSBs.  The spectroscopic observations of host galaxies linked
LSBs preferentially to star-forming (albeit, dwarf) galaxies which
suggested a massive star origin. The twelfth {\it BeppoSAX} event,
GRB\,980425, was very close, 45\,Mpc, and became the prototype for
the class of low luminosity GRBs.  The event was rapidly identified
with a mildly relativistic supernova -- two firsts -- and led to
the induction of the word collapsar into the astronomical lexicon.
Over time, ``red bumps" were seen in optical afterglows of some
``classical" LSBs -- pointing to underlying supernovae.  Pan-chromatic
afterglows provided additional diagnostics: calorimetry, collimation
angle of the relativistic ejecta, circumstellar medium density and
angular size.  Spectral features  in X-ray afterglows were reported
but also, as in the past, did not stand subsequent  scrutiny and
statistics.

\begin{figure}[htbp]
  \centering
 \includegraphics[height=2.7in]{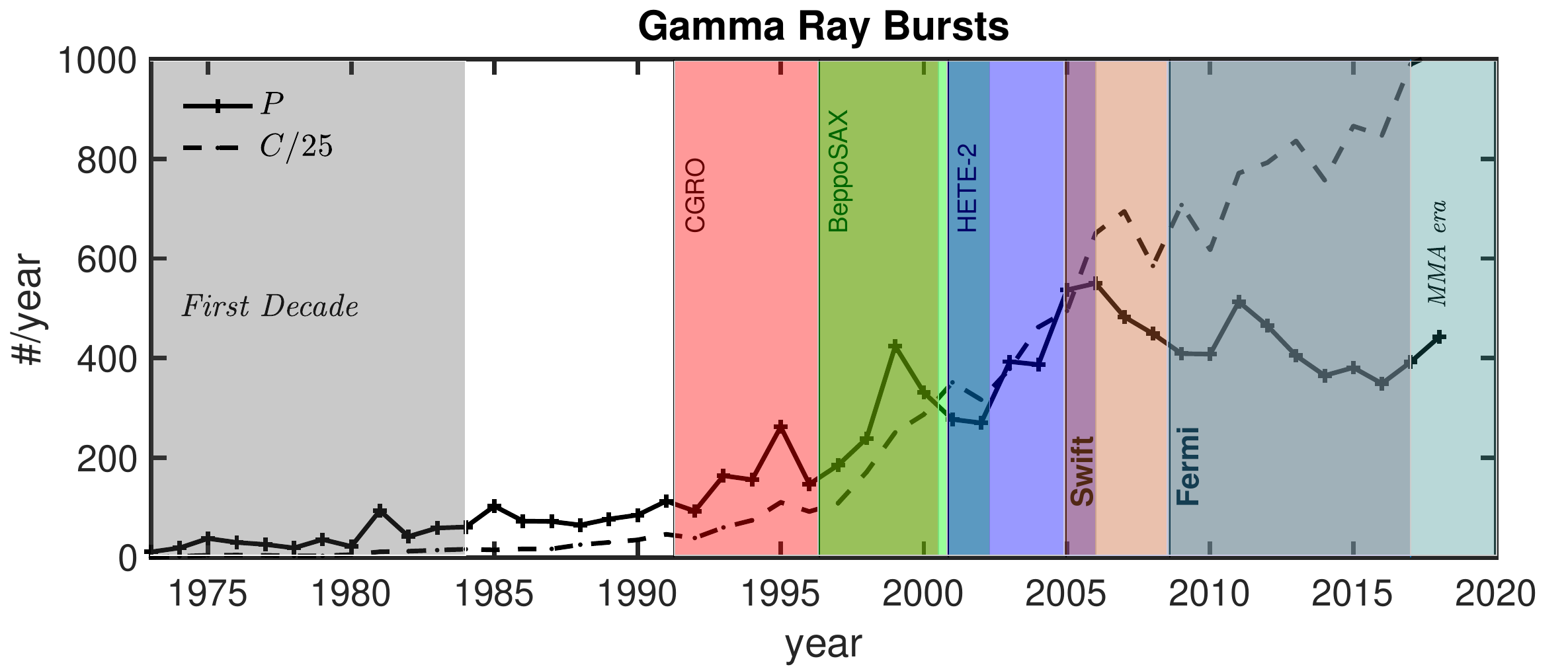}
   \caption[A]{\scriptsize {\bf The annual flux of refereed papers
   ($P$) and  {\it earned} citations ($C$, but divided by 25) of
   papers on or about GRBs since their discovery in 1973.}
   Recognizing that there are 4 more months left in 2018,  $P$ and
   $C$ have been boosted for this year by $3/2$.
   The annual flux of papers is the number of papers
   published in a given year.  In contrast, citations can be sorted
   by the year in which the paper was cited ({\it earned}) or tagged
   to the year in which the paper was published ({\it accrued}).
   Accrued citations, although commonly used, are not useful since
   they lag paper production by (typically) five years.  To obtain
   the bibliographic data I used the  ``modern form" of Astrophysical
   Data System (ADS; \url{http://adsabs.harvard.edu/}) with the
   filter set to ``GRB" or ``gamma ray burst" (also $\gamma$-ray
   burst) in the title or abstract but with ``solar", ``blazar" and
   ``Erratum" excluded.  Over decades, Kevin Hurley of the  {\it
   IPN} painstakingly assembled a list of GRB papers
   (\url{http://www.ssl.berkeley.edu/ipn3/bibliogr.html}).  Hurley's
   list stands at 13,209 but includes conference proceedings and
   reports.  My list is strictly limited to refereed papers and,
   as of September 2018, it stands at 10,374  papers with 388,036
   citations.  The left-most shaded region corresponds to the first
   decade and the right most to the burgeoning ``Multi-Messenger
   Astronomy" (MMA) era.  During the first decade the Soviet Venera
   missions made key contributions (isotropical distribution and
   the discovery of soft gamma-ray repeaters). The IPN was key to
   arcminute localization.  Thin vertical lines mark the launch of
   key GRB missions (which are noted to the right of each line).
   {\it CGRO},  launched in April 1991, was the second of the four
   elements of the Great Observatories program.   {\it BeppoSAX},
   an Italian-Dutch mission, stands for  {\it S}atellite per {\it
   A}stronomia a raggi {\it X} ({\it SAX}) and was named in honor
   of Giuseppe ``Beppo" Occhialini who was the co-discoverer of the
   pion. The High Energy Transient Explorer-2 ({\it HETE-2}) was a
   small mission led out of the Massachusetts Institute of Technology.
   These three missions were terminated or ceased to operate in
   2000, 2002 and 2006, respectively.  The Neil Gehrels {\it Swift}
   Observatory, a ``medium" explorer devoted to studies of GRBs,
   was launched in November, 2004. The {\it Fermi} $\gamma$-ray
   Space Telescope is a major NASA-DOE project aimed at studying
   the $\gamma$-ray sky.  Both {\it Swift} and {\it Fermi} are still
   in operation.
 }
  \label{fig:GRB_Cites}
\end{figure}

The period May--July of 2005 was magical.  {\it Swift} found the
first X-ray afterglow of an SHB. Thereafter, {\it HETE-2} and  {\it
Swift} successively localized two SHBs. Radio and optical afterglows
were discovered. However, unlike LSBs, there were no associated
bright supernovae.  For one event, the host  was a star-forming
galaxy but an elliptical for the other!  {\it Swift} then went on
to accumulate samples of SHB host galaxies (including offsets) and
followup observations for one SHB resulted in a tentative identification
of a long-lived kilonova.  Over time, the sub-classes of ultra-long
duration GRBs and events with the bulk of their energy in the soft
X-ray band (``X-ray flashes") have emerged.  Searches for ``dirty
fireballs" (with peak energy in the ultra-violet or UV band) are
continuing.  Nonetheless, the declining production of papers shows
the maturation of the GRB field (Figure~\ref{fig:GRB_Cites}).

%Kevin's table has an entry line for each year 
 %(eg. 1972 is line 1, before any paper is entered).
% So subtract 2018-1971=47 entries!

It is time to review the lessons learnt from the development of GRB
astronomy.  The first lesson is, excluding solar bursts,  that there
are at least four  {\it major} GRB phenomena: terrestrial gamma-ray
flashes (from lightning in our atmosphere); SGRs, flares from
strongly magnetized neutron stars, readily detected in our Galaxy;
SHBs, coalescence of neutron stars, with typical redshift, $z\approx
0.8$;  LSBs, deaths of a certain select type of massive stars, and
can be seen to the edge of the Universe. The Galactic model was
hobbled by fruitlessly trying to unify SGRs and GRBs.

\begin{figure}[htbp]
 \centering \includegraphics[width=5.5in]{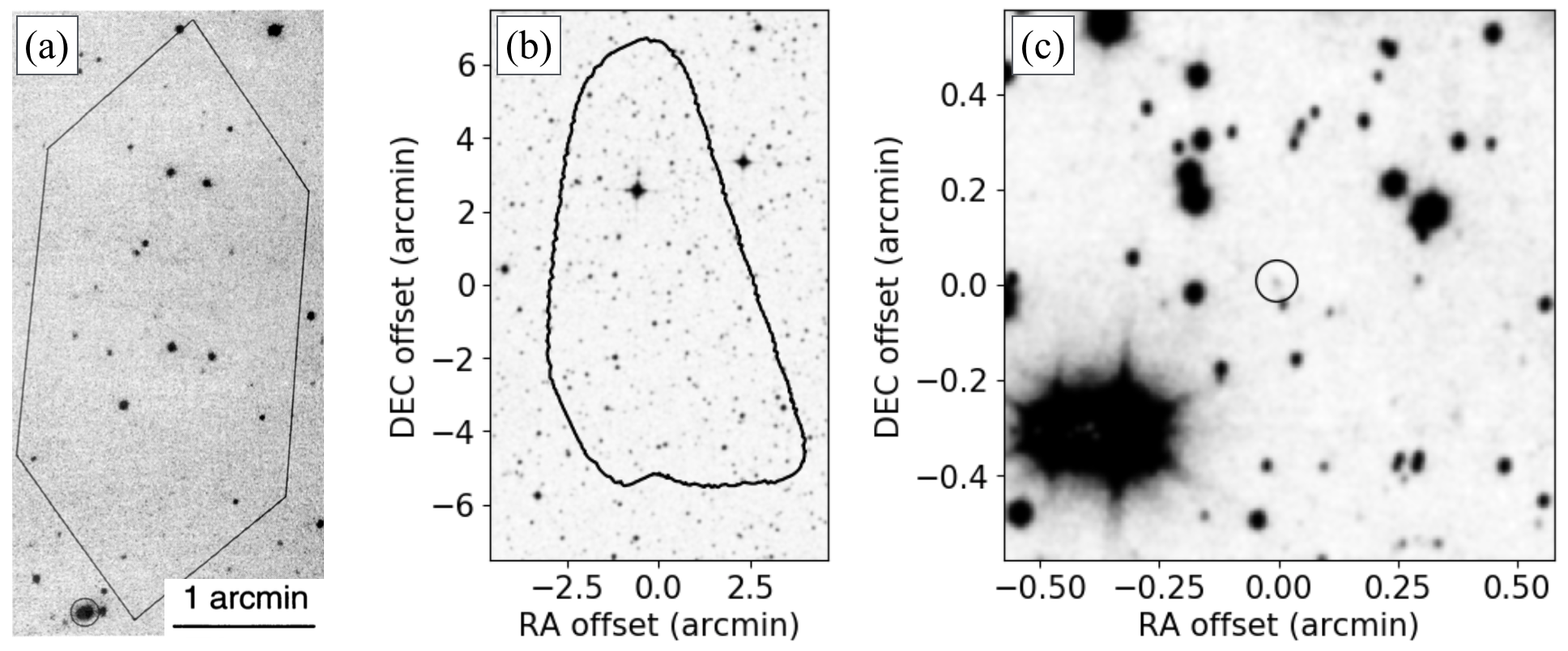}
  \caption{\scriptsize
   {\bf A selection of GRB and FRB localization regions.} Panel
  ({\bf a}) shows the {\it IPN} 99\% confidence error box of
  GRB\,910219 plotted on a deep Keck/NIRC K-band near-IR image.
  Panel ({\bf b}) shows the 99\% localization region of the ``Lorimer''
  burst (FRB\,010724; ref.\ 2) obtained by comparing its detections
  in multiple beams of the Parkes multi-beam receiver.  The background
  image is a composite of Second Digitized Sky Survey red and blue
  plates.  Panel ({\bf c}) shows a deep Keck/LRIS R-band image of
  the field of the Arecibo Repeater (FRB\,121102; ref.\ 13). The
  Repeater has been localized by the Jansky Very Large Array  to a
  dwarf galaxy (marked by a circle) remarkable only for its strong
  emission lines.  Thanks to the astounding precision of Very Long
  Baseline Interferometry, it has been found that FRB\,121102 is,
  to within milli-arcseconds, coincident with a strong non-thermal
  source which has been suggested to be an active galactic nucleus
  or a plerion.  Credit: panel {\bf a} reproduced from ref.\ 14,
  AAS.
   }
 \label{fig:HostCollage}
\end{figure}

Spectral lines top my list of red herrings. When two harmonically
related absorption lines are reported it is difficult {\it not} to
think of cyclotron lines and thence neutron star surfaces.   Next,
early {\it IPN} error boxes (dominated by LSBs) did not show bright
galaxies.  The huge range of galaxy luminosities compounded with
the equally broad $\gamma$-ray luminosity function  and the rapid
redshift evolution of dwarf star-forming galaxies  greatly weakened
this clue (see Figure~\ref{fig:HostCollage}).  Finally, relativistic
beaming results in under-estimating the distance (for a fixed
isotropic energy) and hence over-estimating the host galaxy brightness.

It is said that it is tough to make predictions, especially about
the future. However, the past, objectively analyzed, may improve
our prediction of the future. The key lesson from GRBs is that the
lack of a distance scale led to decades of speculations with little
of lasting value. The great revolution in GRB astronomy can be
directly linked to localization. The first facility which can
routinely localize  ``classical" FRBs will become the {\it BeppoSAX}
of the FRB field!  Next, some clues carry more weight than others.
The isotropy of faint events should have led to an immediate
abandonment of the Galactic model.  Finally, exceptional events
(usually the nearest), if caught, are usually pivotal. {\it HETE-2},
a ``low-cost explorer",  was up when GRB\,030329, one of the closest
LSB exploded and thereby definitively connected LSBs to supernovae.

\begin{figure}[tbp]
  \centering
   \includegraphics[width=4in]{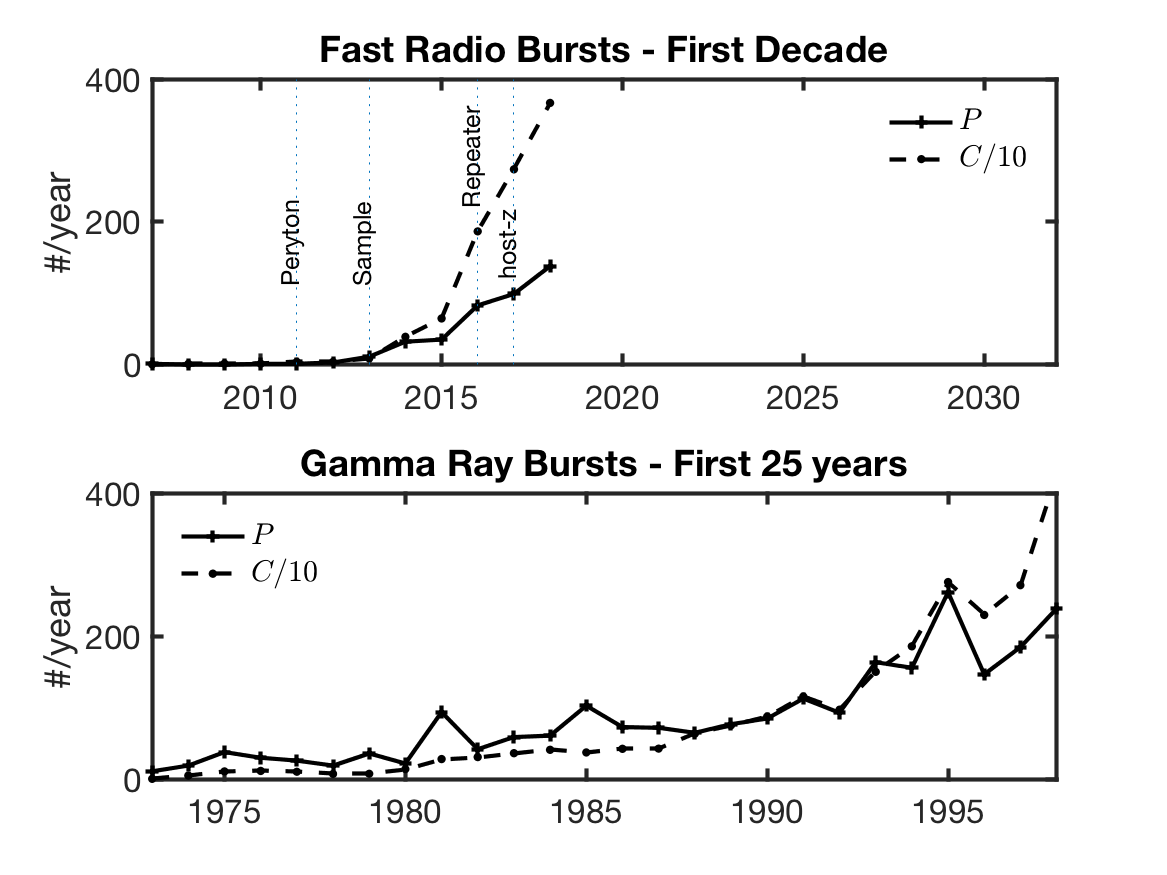}
    \caption{\scriptsize
   The annual flux of refereed papers, $P$ and citations, $C$ (but
   divided by 10) of FRB papers (top) and GRB papers (bottom). As
   in Figure~\ref{fig:GRB_Cites}, $C$ and $P$ for FRB papers in
   2018 were increased by $3/2$.  I followed a similar routine
   described in Figure~\ref{fig:GRB_Cites}: ADS was queried for
   papers with FRB or ``fast radio burst" in the title or abstract.
   I then reviewed each entry and culled unrelated papers.  Conference
   papers were ruthlessly excluded.  Given the youth of this field,
   papers that were submitted but not yet published were included.
   Over the period 2007--September 2018 the FRB field produced 358
   papers and racked up 8224 citations.  Over a similar interval
   the GRB field had 396 papers but only 1630 citations. Clearly
   the FRB field has a higher tempo than that of GRBs.  It took two
   decades for the GRB field to reach the citations that the FRB
   field achieved in its first decade.  The contraction of timescales
   can be reasonably attributed to a larger pool of astronomers.
   For instance, membership of the American Astronomical Society
   increased from 3500 in 1980 to 7600 this year.  The vertical
   lines mark key developments.  ``Perytons" shared some features
   with FRBs and cast grave doubts about the extra-terrestrial
   nature of FRBs.  Perytons were eventually localized to the
   Visitor's Centre of  the Parkes Observatory. In retrospect
   Perytons are the equivalent of (photographic) ``plate defects"
   (or, if we are lucky, the equivalent of spectral lines) of the
   GRB era. The FRB field really took off after the publication of
   a sample of four FRBs.  The localization of the Arecibo Repeater
   and the subsequent identification of its host galaxy at $z\approx
   0.19$ further increased the tempo of the field. Future progress
   will primarily be governed  bythe rate of localizations.  Over
   this and coming years several localization facilities are expected
   to come on line: ASKAP, TRAPUM/MeerKAT, UTMOST2D and DSA-110.
}
 \label{fig:FRB_Cites}
\end{figure}

% "Federal Funding of Astronomical Research", National Academy Press 
 %Figure 2.1 There were 4000 astronomers in 1984 and 7000 by  1999

From Figure~\ref{fig:FRB_Cites} we see that the FRB field, measured
by citations, has moved at more than twice the speed of the GRB
field.  As an example, I note that two years ago a decaying radio
source was linked to an FRB. This claim was investigated ferociously
and persuasive arguments against were published within a month!
This occasion made me wonder where the equivalent of spectral lines
of the GRB era are lurking in the FRB field.

Based on the history of GRB light curves and radio pulsars, meaningful
insights into the nature of FRB sources (let alone the emission
mechanisms) are unlikely to come from details of the light curves.
On the other hand, clues that arise from simpler phenomena --
free-free absorption, rotation measures and multi-path propagation
(scattering, scintillation) -- will likely be powerful. CHIME with
its large collecting area is superbly suited for this purpose.

With no doubt most of the FRBs are {\it not} in our Galaxy.  Next,
there are now two classes:  (the so far sole) Arecibo Repeater and
``classic" FRBs.  The Repeater has been localized to a $z=0.19$
galaxy (Figure~\ref{fig:HostCollage}).  The nature of host galaxies
identified in the first phase of localization -- star-forming galaxy,
elliptical galaxy, intergalactic medium -- will provide robust clues
to the origin of FRBs.  It will, however, require arc-second
localization (see Figure~\ref{fig:HostCollage}).  The offset of the
events with respect to the galaxy components -- nuclear, disk and
halo-- constitute the next phase but will need  {\it sub-arcsecond}
localization. Fortunately, an armada of radio localization facilities
will be coming on line soon (Figure~\ref{fig:FRB_Cites}).  FRB
leaders, informed by the history of GRBs and Gravitational Wave
astronomy, should take note of the vital importance of prompt
dissemination of localization.  Separately, inspired by SGRs, it
would be worthwhile to look for Galactic versions of FRBs.

In preparing for this article I experienced {\it deja vu} all over
again.  The modus operandi was astonishingly familiar: suggest all
possible collisions between comets, asteroids, brown dwarfs, white
dwarfs, neutron stars and black holes, invoke improbable scenarios
(white dwarf and intermediate mass black hole in an eccentric orbit)
and failing all these possibilities invoke string theory.  The
expected diversity of FRB classes should not be used as a license
to speculate recklessly.  A worthwhile  model does not violate basic
physical principles, identifies a mechanism which produces the
observed  signal strength and has an observationally meaningful
rate.  FRBs opened our eyes to extra-galactic millisecond bursts.
In this spirt, searches for Fast Optical Bursts should be undertaken.
Searches commensal with FRB surveys enjoy natural immunity against
local errors (cf.\ the ``plate defects" in the GRB era).  I note
that modern optical time domain surveys are finding$^{12}$ afterglows
even without  GRB triggers!

The greatest impact of GRBs has been in stellar astronomy. LSBs
have shown that some massive stars die with a hyper-relativistic
explosion.  Next, just over a year ago, {\it Fermi} discovered and
promptly reported a faint short hard $\gamma$-ray burst. The burst
was delayed two seconds relative to the merger of two neutron stars,
the celebrated gravitational wave source GW\,170817. This combination
inaugurated the MMA era.  A mission in a high orbit,  perhaps named
``{\it G}amma ray {\it U}ltr{\it A} {\it V}iolet {\it A}stronomy
{\it SAT}ellite ({\it GUAVASAT!}) carrying omni-directional
$\gamma$-ray detectors tuned to SHBs and a UV wide-field imager
would be a strategic space-based mission for this era.  {\it GUAVASAT}
would  also lift the declining fortunes of the GRB field
(Figure~\ref{fig:GRB_Cites}).

The amazing diagnostics of FRBs --  electron column density, magnetic
field and turbulence -- will most assuredly  advance our understanding
of the inter- and circum-galactic medium and perhaps even contribute
to cosmography.  Should it turn out that a class of FRBs is related
to other phenomena (e.g.\ coalescence of neutron stars) or is truly
of exotic origin (e.g. cosmic strings) then the FRB field will
continue to flourish well into its third decade.

\bigskip
\bigskip

\noindent
{\small
I  thank Edwin Henneken, ADS/Center for Astrophysics, for help with
ADS programming and Vikram Ravi for providing Figure~\ref{fig:HostCollage}.
I gratefully acknowledge useful and enlightening discussions with
M.\ Bailes, S.\ B. Cenko, D.\  A.\ Frail,  D.\ A.\ Perley,  E.\ S.\
Phinney \&\ V.\ Ravi. I thank K.\ Plant for careful reading and
Caltech librarian J.\ Painter for help with in finding old references.
}

\bigskip
\bigskip

\noindent {\bf REFERENCES}
\medskip
{\scriptsize
\par\noindent 1. Klebesadel, R.\ W., Strong, I.\ B. and Olson, R.\ A.
Observations of Gamma-Ray Bursts of Cosmic Origin.
{\it Astrophys. J.} {\bf 182}, L85--L88 (1973).   %kso73
\par\noindent 2. Lorimer, D. R., Bailes, M., McLaughlin, M. A. et al.
A Bright Millisecond Radio Burst of Extragalactic Origin.
{\it Science} {\bf 318}, 777--780 (2007).
\par\noindent 3. Ruderman, M. Theories of $\gamma$-ray bursts. 
{\it Annals NY Academy of Sciences} {\bf 262}, 164--180 (1975).   %r75
\par\noindent 4. Ho, C., Epstein, R.\ I.\ and Fenimore, E.\ E.  Gamma-ray bursts. Observations, analyses \&\ theories, Cambridge University Press, 516p. (1992) %hef92
\par\noindent 5. Paczy\'nski, B.  Gamma-ray bursters at cosmological distances.
{\it Astrophys. J.} {\bf 308}, L43-L46 (1986).  %p86
\par\noindent 6.  Fishman, G.\ \& Meegan, C.  Gamma-Ray Bursts.
{\it Annu. Rev. Astron. Astrophys.} {\bf 33}, 415--458 (1995) %fm95
\par\noindent 7. Woosley, S.\ E. Theory of gamma-ray bursts. American Inst. Conf. Series {\bf 384},
709--718 (1997).   %w97
\par\noindent 8. Piran, T. Gamma-ray Bursts and Binary Neutron Star Mergers.
I.\ A.\ U. Symposium 165, 489--502 (1996). %p96
\par\noindent 9. M\'esz\'aros, P.\ \&\ Rees, M.\ J.  Optical and Long-Wavelength Afterglow from Gamma-Ray Bursts.
{\it Astrophys. J.} {\bf 476}, 232--237 (1997).   %mr97
\par\noindent 10. Costa E., Frontera, F., Heise, J., Feroci, M., in't Zand, J. et al. 
Discovery of an X-ray afterglow associated with the $\gamma$-ray burst of 28 February 1997.
{\it Nature} {\bf 387}, 783--785  (1997).  %cfh+97
\par\noindent 11. Metzger, M.\ R., Djorgovski, S.\ G., Kulkarni, S.\ R., Steidel, C.\ C., Adelberger, K.\ L.\ et al. Spectral constraints on the redshift of the optical counterpart to the {$\gamma$}-ray burst of 8 May 1997.
{\it Nature} {\bf 387}, 878--880 (1997)   %mdk+97
\par\noindent
12. Cenko, S. B., Urban, A. L, Perley, D. A.  et al.
iPTF14yb: The First Discovery of a Gamma-Ray Burst Afterglow Independent of a High-energy Trigger.
 {\it Astrophys. J.} {\bf 803}, L24--L28 (2015).
\par\noindent
13.
Spitler, L.\ G.\, Scholz, P., Hessels, J.\ W.\ T.\ et al.
A repeating fast radio burst.
{\it Nature} {\bf 531}, 202--205 (2016)
\par\noindent 
14. Larson, S.\ B., McLean, I.\ S., Becklin, E.\ E. Luminous Galaxies near Gamma-Ray Burst Positions.
{\it Astrophys. J.} {\bf  460}, L95--L97 (1996) %lmb96

}
\end{document}